\documentclass[12pt]{article}
\usepackage{axodraw,epsfig}
\usepackage{graphics}
\usepackage{color}

\usepackage{amsmath,amssymb}
\newcommand{\bea}{\begin{eqnarray}}
\newcommand{\beas}{\begin{eqnarray*}}
\newcommand{\nn}{\nonumber}
\newcommand{\eea}{\end{eqnarray}}
\newcommand{\eeas}{\end{eqnarray*}}
\newcommand{\bd}{\begin{displaymath}}
\newcommand{\ed}{\end{displaymath}}
\newcommand{\be}{\begin{equation}}
\newcommand{\ee}{\end{equation}}
\newcommand{\bi}{\begin{itemize}}
\newcommand{\ei}{\end{itemize}}

\newcommand{\sa}{s_{\alpha}}
\newcommand{\ca}{c_{\alpha}}

\newcommand{\ep}{\epsilon}
\newcommand{\tep}{\tilde{\epsilon}}

\newcommand{\tb}{\tan \beta}
\newcommand{\tbq}{\tan^2 \beta}
\newcommand{\tbt}{\tan^3 \beta}

\newcommand{\as}{\alpha_s}

\newcommand{\hc}{\mbox{h.c.}}

\newcommand{\Br}{{\rm Br}}

\catcode`@=11
\newcount\@tempcntc
\def\@citex[#1]#2{\if@filesw\immediate\write\@auxout{\string\citation{#2}}\fi
  \@tempcnta\z@\@tempcntb\m@ne\def\@citea{}\@cite{\@for\@citeb:=#2\do
    {\@ifundefined
       {b@\@citeb}{\@citeo\@tempcntb\m@ne\@citea\def\@citea{,}{\bf ?}\@warning
       {Citation `\@citeb' on page \thepage \space undefined}}%
    {\setbox\z@\hbox{\global\@tempcntc0\csname b@\@citeb\endcsname\relax}%
     \ifnum\@tempcntc=\z@ \@citeo\@tempcntb\m@ne
       \@citea\def\@citea{,}\hbox{\csname b@\@citeb\endcsname}%
     \else
      \advance\@tempcntb\@ne
      \ifnum\@tempcntb=\@tempcntc
      \else\advance\@tempcntb\m@ne\@citeo
      \@tempcnta\@tempcntc\@tempcntb\@tempcntc\fi\fi}}\@citeo}{#1}}
\def\@citeo{\ifnum\@tempcnta>\@tempcntb\else\@citea\def\@citea{,}%
  \ifnum\@tempcnta=\@tempcntb\the\@tempcnta\else
   {\advance\@tempcnta\@ne\ifnum\@tempcnta=\@tempcntb \else \def\@citea{--}\fi
    \advance\@tempcnta\m@ne\the\@tempcnta\@citea\the\@tempcntb}\fi\fi}
\catcode`@=12

\begin{document}

\begin{titlepage}

\begin{flushright}
\normalsize
PITHA~09/04\\
CERN--PH--TH/2009-017\\
SFB/CPP-09-09 \\
%0901.xxxx [hep-ph]\\
January 29, 2008\\
\end{flushright}

\vskip1.5cm
\begin{center}
\Large\bf\boldmath
Hadronic $B$ decays in the MSSM with large $\tan\beta$
\unboldmath
\end{center}

\vspace*{0.8cm}
\begin{center}

{\sc M.~Beneke}$^{a,b}$,
{\sc Xin-Qiang~Li}$^{a,}$\footnote{Alexander-von-Humboldt Fellow},
and {\sc L.~Vernazza}$^{a}$\\
\vspace{0.7cm}
{\sl $^a$
Institut f\"ur Theoretische Physik E,
RWTH Aachen University,\\
D--52056 Aachen, Germany}\\[0.2cm]
{\sl $^b$
CERN, Theory Division,\\
CH -- 1211 Gen\`{e}ve, Switzerland}\\[0.2cm]
\end{center}

\vspace*{0.8cm}
\begin{abstract}
\noindent
We present an analysis of non-leptonic $B$ decays in the
minimally flavour-violating MSSM with
large $\tan \beta$. We relate the Wilson coefficients of the relevant
hadronic scalar operators to leptonic
observables, showing that the present limits on the $B_s \to \mu^+\mu^-$
and $B^+\to \tau^+ \nu_{\tau}$ branching fractions exclude any visible
effect in hadronic decays. We study the transverse helicity amplitudes
of $B \to VV$ decays, which exhibit an enhanced
sensitivity to the scalar operators, showing that even though
an order one modification relative to the SM is not excluded in some
of these amplitudes, they are too small to be detected at
$B$ factories.

\vspace*{0.8cm}
%\noindent
%PACS numbers: 12.38.Bx, 14.40.Gx, 14.65.Fy, 14.65.Ha

\end{abstract}
\vfil
\end{titlepage}

\newpage

\section{Introduction}
\label{intro}

If new particles exist at the TeV scale, then the striking absence
of evidence so far for their virtual effects in $B$ or $K$ meson mixing
and decay suggests that the pattern of flavour-changing interactions
is governed by the standard-model (SM) Yukawa coupling matrices even at the
TeV scale. The minimal supersymmetric SM (MSSM) with large ratio
$\tan\beta$ of the Higgs vacuum expectation values and no new sources
of flavour violation in the supersymmetry-breaking Lagrangian is an example
of such a minimally flavour-violating (MFV) theory, which nevertheless
may exhibit sizeable differences from the SM due to Higgs
exchange. The leptonic $B_s\to \mu^+\mu^-$ and $B^+\to \tau^+ \nu_\tau$
decays have been extensively studied in this model, as well as meson mixing
and $B\to D \tau\nu_\tau$ \cite{Hamzaoui:1998nu,Babu:1999hn,Chankowski:2000ng,Huang:2000sm,Bobeth:2001sq,Degrassi:2000qf,Carena:2000uj,Isidori:2001fv,Buras:2002wq,Buras:2002vd,Akeroyd:2003zr,Isidori:2006pk,Nierste:2008qe}.
Higgs exchange also generates scalar four-quark operators, which contribute
to non-leptonic $B$ decays. The effects of scalar operators on non-leptonic
$B$ decays have been studied in the MSSM (not necessarily minimally
flavour-violating) and a general two Higgs doublet
model in~\cite{Cheng:2002ay,Cheng:2003im,Das:2004hq,Huang:2005qb,Cheng:2006qu,Wu:2007eb,Hatanaka:2007mp}, mostly
in connection with transverse polarization in $B$ decays to two
vector mesons (VV), and for specific decay modes. Some of these
studies find large deviations from SM expectations for non-leptonic
decays.

The present work is motivated by the question whether, given the present
strong constraints from the leptonic decays, further insight
on the MFV MSSM at large $\tan\beta$ can be derived from
charmless non-leptonic
$B$~decays. To this end, extending previous analyses, we relate directly
the Wilson coefficients of the leptonic to the relevant
hadronic scalar operators, including charged Higgs
exchange effects, and calculate the hadronic matrix elements in QCD
factorization~\cite{Beneke:1999br,Beneke:2001ev}. We also study observables
related to the helicity amplitudes of $B\to VV$, which exhibit an
enhanced sensitivity to the Higgs-induced scalar operators. We find
that the present limit on the $B_s\to \mu^+\mu^-$ branching fraction,
and the observation of $B^+\to \tau^+ \nu_\tau$ with a branching fraction
close to the SM expectation, exclude any
visible effects in hadronic decays, but for an academic exception:
the positive-helicity amplitude of $\bar B\to VV$ modes may receive
order one modifications relative to the SM. However, this amplitude is
too small to be detected at present or planned $B$ factories.

\section{\boldmath
Scalar four-quark operators in the
MSSM with large $\tan\beta$}
\label{weak}

In the SM the effective Hamiltonian for charmless $B$ decays
is
\be
{\cal H}_{\rm eff}^{\rm SM} = \frac{G_F}{\sqrt{2}}
\sum_{p=u,c} \lambda_{p}^{(D)} \left(C_1
Q_1^p+C_2Q_2^p+\sum_{i=3}^{10} C_i Q_i
+C_{7\gamma}Q_{7\gamma}+C_{8g}Q_{8g}\right)+\hc,
\label{a1}
\ee
where  $D=d$ or $s$ depending on the decay mode considered, and
$\lambda_p^{(D)}=V_{pb}V_{pD}^*$ denotes a product of CKM matrix
elements. The conventions for the operators $Q_i$ and the
approximations for the short-distance coefficients $C_i$ are
given in \cite{Beneke:2001ev}. Here we only note that the four-quark
``current-current'' and ``penguin'' operators
$Q^p_{1,2},Q_{3-10}$ are all of the
$(V-A) \times (V\mp A)$ form.

\begin{figure}[t]
  \vskip-1.0cm
  \begin{center}
  \hspace*{-4.0cm}
  \includegraphics[width=1.4\textwidth]{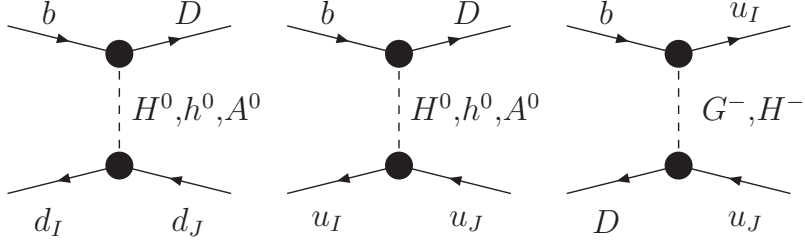}
  \vspace*{-27cm}
  \caption{Four-quark interactions mediated by neutral and charged Higgs bosons.}
  \label{fig1}
  \end{center}
\end{figure}

In the MSSM new four-quark operators are generated and the coefficients
of the SM operators are modified. We consider the large-$\tan\beta$
scenario in a set-up, where the superpartner particles
are somewhat heavier than the electroweak gauge bosons and the Higgs
bosons (the ``decoupling limit''), such that the leading effect is due to
Higgs exchange not only for the neutral but also for the charged current
interactions, as
shown in figure~\ref{fig1}. Of particular interest are the flavour-changing
neutral Higgs couplings to fermions, which originate from a loop-induced
coupling of the ``wrong'' Higgs field $H_u$ to the down-type quarks,  since
these couplings are enhanced by several powers
of $\tan\beta$~\cite{Babu:1999hn,Isidori:2001fv,Buras:2002vd}. In the
following we use the effective couplings given
in~\cite{Isidori:2001fv,Buras:2002vd} in the decoupling limit
to obtain the short-distance
coefficients of the scalar four-quark operators from tree-level
Higgs exchange. The coefficients are then evolved from the electroweak
scale to the bottom mass scale
$m_b$ by the renormalization group equations. The relevant
Higgs-induced terms in the effective Hamiltonian can be written as
\be
{\cal H}_{\rm eff}^{\rm Higgs} = \frac{G_F}{\sqrt{2}}
\sum_{p=u,c} \lambda_{p}^{(D)} \left(C_{11}^D
Q_{11}^p+C_{12}^D Q_{12}^p+\sum_{i=13}^{14} \,\sum_{q=d,s,b} C_i^q Q_i^q
\right)+\hc,
\label{a2}
\ee
similar to (\ref{a1}). The ``current-current'' operators
\be
Q_{11}^p = (\bar p_i b_i)_{S+P} \,(\bar D_j p_j)_{S-P},
\qquad
Q_{12}^p = (\bar p_i b_j)_{S+P} \,(\bar D_j p_i)_{S-P},
\qquad (p=u,c)
\label{scalartree}
\ee
originate from charged Higgs exchange; the ``penguin'' operators
\be
Q_{13}^q = (\bar D_i b_i)_{S+P} \,(\bar q_j q_j)_{S-P},
\qquad
Q_{14}^q = (\bar D_i b_j)_{S+P} \,(\bar q_j q_i)_{S-P},
\qquad (q=d,s,b)
\label{scalarpenguin}
\ee
from the loop-induced neutral Higgs-fermion vertices. Here
$i,j$ denote colour indices and $(\bar q q)_{S\pm P} =
\bar q\,(1\pm \gamma_5)q$. The CKM factors $\lambda_{p}^{(D)}$
in (\ref{a1}), (\ref{a2}) are now assumed to be composed of the effective
CKM matrix elements $V_{ij}^{\rm eff}$ that correspond to the low-energy
couplings.

\subsubsection*{\boldmath Neutral Higgs exchange: $b\to D \bar q q$ transitions}

It is straightforward to assemble the short-distance coefficients from
tree-level Higgs exchange in terms of the effective neutral Higgs couplings
given in~\cite{Isidori:2001fv,Buras:2002vd}. Combining a flavour-changing
and a flavour-conserving coupling, we find in the large-$\tan\beta$
limit, where $\sin\beta\approx 1$, $1/\cos\beta\approx \tan\beta$:
\bea
C_{13}^{d_J}(\mu_{H}) &=&
\frac{1}{2} \frac{\bar{m}_{d_J} \bar{m}_b \ep_Y y_t^2 \tbt}
{(1+\tep_3 \tb)(1+\ep_0 \tb)(1+\tep_J \tb)} \,\mathcal{F}_{2,J}^-,
\qquad
C_{14}^{d_J}(\mu_{H})=0.
\qquad
\label{c1314}
\eea
Here
\be
\mathcal{F}_{2,J}^- = \frac{s_{\alpha-\beta}\left(\ca+\tep_J
\sa\right)}{M^2_{H^0}} +\frac{c_{\alpha-\beta}\left(-\sa+\tep_J
\ca\right)}{M^2_{h^0}} - \frac{1}{M^2_{A^0}}
\approx -\frac{2}{M_{A^0}^2},
\label{F2minus}
\ee
with $\ca \equiv \cos \alpha, \ldots$. The
$\epsilon$-coefficients appearing in (\ref{c1314}) are defined
in \cite{Buras:2002vd} and denote the loop-induced Higgs-fermion
couplings. In the large-$\tan\beta$ MSSM products $\epsilon\times
\tan\beta$ can be of order one. Just as in the
$b\to D \ell_J \ell_J$ transitions, the coefficients of the hadronic
Higgs penguin operators are strongly enhanced by the factor
$\tan^3\beta$. The quark masses $\bar m_q$ are the $\overline{\rm MS}$
masses in the low-energy effective theory at the matching scale
$\mu_H$.

Higgs exchange generates $(\bar D b)_{S-P} \,(\bar q q)_{S+P}$
operators as well, but in this case
the factor $\bar m_b$ in (\ref{c1314}) is replaced by $\bar m_D$, which is
at most $\bar m_s$. The remaining two helicity combinations have short-distance
coefficients multiplied by a function $\mathcal{F}_{2,J}^+$, which
vanishes in the large-$\tan\beta$ limit. Thus, it is sufficient
to consider the operators $Q_{13,14}^q$. The neutral Higgs
coupling to up-type quarks (second diagram in figure~\ref{fig1})
is suppressed at large $\tan\beta$ relative to the down-type quarks,
thus $q=d_J=d,s,b$. In fact, the operators $Q_{13,14}^d$ might also
be dropped due to the small down-quark mass. The operator
$Q_{13,14}^b$ has the largest coefficient, but it contributes
to non-leptonic decays only through loops. Finally, we note
that the double Higgs penguin diagrams (first diagram in figure~\ref{fig1}
with flavour change at both vertices) are irrelevant to non-leptonic
decays due to their extra CKM suppression. We therefore conclude
that in the  MFV MSSM with large $\tan\beta$, only a small set of
scalar penguin operators $Q_{13,14}^{s,b}$ is relevant. Of these
$Q_{14}^{s,b}$ is absent at tree-level, but it is kept for the moment, since
it may be generated by renormalization group evolution
(see discussion below).

\subsubsection*{Charged Higgs exchange}

The operators $Q_{11,12}^p$ arise from the third diagram in figure
\ref{fig1}. Once again only the $(S+P)\times (S-P)$ Dirac structure
is dominant at large $\tan\beta$. For charmless decays we need
only the cases $u_I=u_J=p=u,c$, and obtain
\be
C_{11}^D(\mu_{H}) =
-\frac{\bar{m}_{b}\bar{m}_{D}}{M_{H^+}^2} \frac{\tbq}{(1+\epsilon_0
\tb)^2},
\qquad
C_{12}^D(\mu_{H}) = 0.
\label{c1112}
\ee
Although $C_{11}^D$ is enhanced only by $\tbq$, there is no loop
suppression factor $\ep_Y$. Due to the factor $\bar m_D$, charged
Higgs exchange is relevant in practice only for $b\to s$ transitions.

\subsubsection*{Renormalization group evolution}
\label{RGE}

We first discuss the evolution of the short-distance coefficients from
a typical Higgs mass scale, which we assume to be $\mu_H=200\,$GeV,
to the bottom mass scale
$m_b=4.2\,$GeV, when penguin diagrams are neglected. Then each pair
of operators $(Q_{11}^p,Q_{12}^p)$, $(Q_{13}^q,Q_{14}^q)$ evolves
independently in leading logarithmic (LL) accuracy with anomalous
dimension matrix (in units of $\alpha_s/(4\pi)$)
\be
\label{a31}
{\gamma_{2 \times 2}}=
\left(
\begin{array}{ccc}
  -16 & & 0 \\
   -6 & & 2 \\
\end{array}
\right).
\ee
With $\as(m_b)/\as(\mu_H) \approx 2.13$, this results in
\be
C_{11}^D(m_b)/C_{11}^D(\mu_H)=C_{13}^q(m_b)/C_{13}^q(\mu_H) \approx 2.20,
\label{evolvecs}
\ee
while $C_{12}^D(\mu)$, $C_{14}^q(\mu)$ remain zero. Using the 2-loop
NDR scheme anomalous dimension matrix (ADM)~\cite{Buras:2000if}
we obtain 2.35 instead and
$C_{12}^D(m_b)/C_{11}^D(\mu_H)=C_{14}^q(m_b)/C_{13}^q(\mu_H)
\approx 0.088$,
but since we do not have the 1-loop correction to
the initial condition of the scalar operators at $\mu_H$, the
next-to-leading logarithmic (NLL)
evolution is not fully consistent. In any case, we conclude that
the operators $Q_{12}^p,Q_{14}^q$ can be neglected to first approximation,
since their coefficient functions are suppressed by a factor 25.

Including penguin diagrams requires to enlarge the operator basis, since
the scalar operators mix at the LL level into the SM penguin operators
as well as their ``mirror'' copies, defined by a global exchange of left- and
right chiralities of the quark fields. For the following discussion
we neglect the electroweak penguin operators, so we deal with the
six SM operators $Q_{1,2}^p$, $Q_{3-6}$, their mirror copies
$Q_{1,2}^{\prime\,p}$, $Q_{3-6}^{\prime}$,
and the six scalar operators $Q_{11,12}^{p}$, $Q_{13,14}^{D,b}$. The
structure of the ADM reads
\be
\label{a30}
{\gamma}=\left(%
\begin{array}{llc}
 {\gamma_{6\times 6}} & {0_{6\times 6}} & {0_{6\times 6}}\\
 {0_{6\times 6}} &  {\gamma_{6\times 6}^\prime} & {0_{6\times 6}}\\
 {\gamma_{6 \times 6}^{\rm sc-p}} &
 {\gamma_{6 \times 6}^{\prime \,\rm sc-p}} &
 {\gamma_{6\times 6}^{\rm sc}}
\end{array}%
\right),
\ee
where $\gamma_{6\times 6}=\gamma_{6\times 6}^\prime$ is the
ADM for the SM current-current and QCD penguin operators (equal for the
mirror operators) and $\gamma_{6\times 6}^{\rm sc}$ is a block-diagonal
matrix with three identical $2\times 2$ blocks given by
$\gamma_{2 \times 2}$ in (\ref{a31}): one for $Q_{11,12}^{p}$,
one for $Q_{13,14}^{D}$, depending on the transition, and one
for $Q_{13,14}^{b}$. The matrices ${\gamma_{6 \times 6}^{(\prime)\,\rm sc-p}}$
describe the mixing of the scalar operators into the penguin
operators. We find that $Q_{11,12}^p$ and
$Q_{13,14}^{D}$ mix into the mirror penguin operators, while
only $Q_{13,14}^{b}$ mixes into the SM penguins. Thus
$[\gamma_{6 \times6}^{\rm sc-p}]^T = (0|0|\Gamma^T)$ and
$[\gamma_{6 \times 6}^{\prime\,\rm sc-p}]^T = (\Gamma^T|\Gamma^T|0)$,
where
\be
\label{a32}
{\Gamma}=
\left(%
\begin{array}{cccccc}
  0 & 0 & \frac{1}{9} & -\frac{1}{3} & \frac{1}{9} & -\frac{1}{3} \\
  0 & 0 & 0 & 0 & 0 & 0 \\
\end{array}%
\right).
\ee
Solving the RGE equations leaves (\ref{evolvecs}) unchanged, generates
the mirror QCD penguin operators with coefficient functions
\be
C_i^{\prime\,D}(m_b) \approx -0.71\, C_i^{\rm SM}(m_b) \times
[C_{11}^D(\mu_H)+C_{13}^D(\mu_H)],
\qquad i=3 \ldots 6,
\label{pen1}
\ee
and modifies the SM penguin coefficients according to
$C_i = C_i^{\rm SM}+\delta C_i$ with
\be
\delta C_i(m_b)
\approx -0.71\, C_i^{\rm SM}(m_b) \times C_{13}^b(\mu_H),
\qquad i=3 \ldots 6.
\label{pen2}
\ee
Since the SM penguin coefficients $C_i^{\rm SM}(m_b)$ are small numbers,
the penguin-mixing effects are small, unless the coefficient functions
of the scalar operators are of order one. However, due to their different
chiral structure, the mirror penguin operators contribute differently
from the standard ones to the transverse helicity amplitudes in
$B\to VV$ decays as discussed below.

\section{\boldmath Constraints from
$B_s\rightarrow\mu^+\mu^-$ and  $B^+\rightarrow\tau^+\nu_\tau$}
\label{Bsmumu}

The natural size of the loop-induced neutral Higgs couplings
$\epsilon_{0},\epsilon_Y,\tilde \epsilon_J$ is of order $0.01$, the precise
values depending on MSSM parameters. Assuming $M_{A^0}=200\,$GeV
and $\tan\beta=50$, this allows the scalar penguin
operators to have coefficients of order $C_{13}^s\simeq 0.01$,
$C_{13}^b\simeq 0.5$, which are comparable to SM penguin coefficients.
However, the non-observation of $B_s\rightarrow\mu^+\mu^-$ implies
much stronger limits on the size of the scalar four-quark
operator coefficient functions.

The decay  $B_s\rightarrow\mu^+\mu^-$ proceeds via an interaction
similar to the first diagram of figure~\ref{fig1} except that the
lower legs are replaced by a muon pair. Since the lower vertex is
a tree-level neutral Higgs coupling, the leptonic and hadronic decay
are closely related. For large $\tan\beta$, a single scalar operator
$\left(\bar{D}b\right)_{S+P}\left(\bar{\mu} \mu\right)_{S-P}$,
similar in structure to $Q_{13}^q$, dominates the $B_s\rightarrow\mu^+\mu^-$
decay amplitude. Its coefficient function is given by
\be
\label{a22}
C_{\mu\mu}(\mu_{H}) =-\frac{1}{2}
\frac{\bar{m}_b m_\mu \ep_Y y_t^2 \tbt} {(1+\tep_3
\tb)(1+\ep_0 \tb)} \mathcal{F}_{2l}^-,
\ee
with
\be
\label{a23}
\mathcal{F}_{2l}^-=\frac{s_{\alpha-\beta}(\ca)}{M^2_{H_0}}
+\frac{c_{\alpha-\beta}(-\sa)}{M^2_{h_0}}-\frac{1}{M^2_{A_0}}
\approx \mathcal{F}_{2,J}^-.
\ee
For large $\tan\beta$, and at the level of the present experimental limit,
the SM contribution to the decay amplitude is
negligible, and the branching ratio is given by
\be
\label{a25}
{\rm Br}(B_s\rightarrow \mu^+\mu^-) = \frac{G_F^2 f_{B_s}^2 m_{B_s}^5
\tau_{B_s}}{8 \pi (\bar m_b+\bar m_s)^2}
\left|\lambda_t^{(s)}\right|^2\left|C_{\mu\mu}\right|^2.
\ee
Comparing (\ref{c1314}) to (\ref{a22}), we see that we can eliminate
$C_{\mu\mu}$ in favour of $C_{13}^q$ in the previous equation and turn it
into
\be
\label{a26}
(1+\tep_J \tb)\,|C_{13}^{d_J}(\mu_H)| =
\frac{2\sqrt{2 \pi}(\bar m_b+\bar m_s)(\mu_H)}{G_F
f_{B_s}m_{B_s}^{5/2}\tau_{B_s}^{1/2}\,|\lambda_t^{(s)}|}\,
\frac{\bar m_{d_J}(\mu_H)}{m_{\mu}} \,\Big[{\rm Br}(B_s
\rightarrow \mu^+\mu^-)\Big]^{1/2}.
\ee
The present experimental upper limit on the $B_s
\rightarrow \mu^+\mu^-$ branching fraction is ${\rm Br}(B_s
\rightarrow \mu^+\mu^-)\leq 5.8 \cdot 10^{-8}$ at 95\%~C.L.~\cite{:2007kv}.
Using $f_{B_s}=240\,$MeV, $\bar m_s(2\,\mbox{GeV})=90\,$MeV and
$\bar m_b(\bar m_b)=4.2\,$GeV, and evolving both quark masses
to the common scale $\mu_H=200\,$GeV, we obtain
\be
(1+\epsilon_0 \tb) \,|C_{13}^{s}(\mu_{H})|
\leq 1.4 \cdot 10^{-4},
\qquad
(1+\tilde \epsilon_3 \tb) \,|C_{13}^{b}(\mu_{H})|
\leq 7.9 \cdot 10^{-3}.
\label{a27}
\ee
When $\epsilon_0$ and/or $\tilde \epsilon_3$ are negative, the
coefficient functions can be larger than the values on the right-hand
side. However, the brackets multiplying the coefficient functions
enter the relation between the quark masses and the down-type Yukawa
couplings, and hence $(1+\tilde \epsilon_3 \tb)$ cannot become very
small, if the bottom Yukawa coupling
is to remain perturbative. We allow a factor of three enhancement
of the coefficient functions to be conservative (that is, the brackets
are required to be larger than $1/3$). Including the factor
(\ref{evolvecs}) from evolution to the scale $m_b$ leads  to
\be
|C_{13}^{s}(m_b)|\leq 0.001, \qquad
|C_{13}^{b}(m_b)| \leq 0.05,
\ee
while $C_{13}^{d}(m_b)$ is a factor $\bar{m}_d/\bar{m}_s$ smaller
than $C_{13}^{s}(m_b)$ and therefore negligible.
Thus, the coefficient functions of the hadronic
flavour-changing neutral Higgs penguin
operators are constrained to be
a factor of 10 smaller than the above estimates derived from
$M_{A^0}=200\,$GeV and $\tan\beta=50$.

The short-distance coefficient $C_{11}^{D}(\mu_H)$, arising
from charged Higgs exchange, can be related to
$B^+\rightarrow\tau^+\nu_\tau$ in a similar way. Using
(\ref{c1112}) the ratio \cite{Akeroyd:2003zr,Hou:1992sy}
\be
\label{a27b}
R_{\tau\nu_{\tau}}\equiv
\frac{\Br(B^+\rightarrow \tau^+ \nu_{\tau})_{\rm
MSSM}}{\Br(B^+\rightarrow \tau^+\nu_{\tau})_{\rm SM}}=
\left(1-\frac{m_B^2}{m_{H^+}^2}\frac{\tbq}{1+\ep_0 \tb}\right)^{\!2},
\ee
is expressed in terms of $C_{11}^{D}(\mu_H)$ as
\be\label{a27c}
R_{\tau\nu_{\tau}}=
\left(1+C_{11}^{D}(\mu_H) \,\frac{m_B^2(1+\ep_0\tb)}{\bar m_D(\mu_H)
\bar m_b(\mu_H)}\right)^{\!2}.
\ee
The present average of the Babar and Belle measurements of the branching
fraction is $\Br(B^+\rightarrow \tau^+ \nu_{\tau}) =
(1.51\pm0.33)\cdot 10^{-4}$ \cite{Barberio:2008fa,Aubert:2007xj,:2008ch}.
Employing the central value $|V_{ub}|\,f_{B_d}=7.4 \cdot 10^{-4}\,$GeV
and assigning a conservative $50\%$ uncertainty to the SM prediction
of the branching fraction, the measurement constrains $R_{\tau\nu_{\tau}}$
to lie in the range
\be
\label{a27d}
0.72<R_{\tau\nu_{\tau}}<2.40.
\ee
Concentrating on the case $D=s$ this implies the allowed ranges
\bea
&&-0.012<(1+\ep_0 \tb) \,C_{11}^{s}(\mu_H)<-0.009,
\quad
\nonumber\\
&&-0.001<(1+\ep_0 \tb) \,C_{11}^{s}(\mu_H) <0.003.
\label{a27e}
\eea
The first range corresponds to the situation, where the charged Higgs
contribution is about twice as large as the SM one, and opposite in sign.
Requiring $1+\ep_0 \tb>1/3$ and including the
RG evolution (\ref{evolvecs}) results in
\be
\label{a27f}
-0.08 < C_{11}^{s}(m_b)<-0.06,
\quad\mbox{or}\quad
-0.005 <C_{11}^{s}(m_b) <0.018.
\ee
The constraint from $B^+\rightarrow\tau^+\nu_{\tau}$ on $C_{11}^s$ is not as
stringent as the one from $B_s \rightarrow \mu^+\mu^-$ on $C_{13}^s$,
but one must remember that the charged Higgs contribution to hadronic
charmless decays must compete with the SM tree operators rather than the
penguin operators. In addition, since $|C^s_{11}| \ll 1$,
the contribution (\ref{pen1}) to the mirror
penguin coefficients remains small. These conclusions hold
{\em a fortiori} for $C_{11}^{d}$, which is a factor of
$\bar m_d/\bar m_s$ smaller than  $C_{11}^{s}$.

To conclude this section, we remark that we also performed a
MSSM parameter space scan, calculating explicitly the loop-induced $\epsilon$
parameters subject to the experimental constraints from
$B_s\rightarrow\mu^+\mu^-$, $B^+\rightarrow\tau^+\nu_\tau$,
$B\rightarrow X_s\gamma$, and $\Delta M_{B_{d,s}}$. Here we also included
the subleading scalar operators for $B_s\rightarrow\mu^+\mu^-$,
as well as the exact expressions for
$\mathcal{F}_{2,J}^-,\mathcal{F}_{2l}^-$ and related functions.
The resulting values of the short-distance coefficients $C_{11}^D$,
$C_{13}^q$ are in agreement with the ranges given above.

\section{\boldmath Hadronic matrix elements for $B\to M_1 M_2$}
\label{fact}

To calculate the decay amplitudes of non-leptonic, charmless $B$ decays,
we employ the QCD factorization (QCDF)
framework \cite{Beneke:1999br,Beneke:2001ev}.
We refer to these papers for a discussion of the method and
to~\cite{Beneke:2003zv,Beneke:2006hg} for the definitions and notation
that we adopt below. Let us emphasize that given the constraints on
the coefficient functions, a leading-order treatment, where QCDF
is equivalent to naive factorization~\cite{Bauer:1986bm}, would suffice.
However, it takes little additional effort to include the
first-order radiative corrections.

The matrix element of the effective Hamiltonian is written as
\be
\label{b2}
\langle M_1'M_2'|{\cal{H}_{\rm eff}}
|\bar{B}\rangle=\sum_{p=u,c}
\lambda_{p}^{(D)} \langle
M_1'M_2'|{\cal{T}}_A^p+{\cal{T}}_B^p|\bar{B}\rangle,
\ee
where  ${\cal{T}}_A^p$ account for vertex, penguin and spectator-scattering
terms in the QCDF formula and  ${\cal{T}}_B^p$ parameterizes the
weak annihilation amplitudes. We generalize the expression given
in \cite{Beneke:2003zv} to account for the scalar amplitudes and those
from the mirror QCD penguin operators, such that now
\begin{eqnarray}\label{alphaidef}
   {\cal T}_A^p
   &=& \delta_{pu}\,[\alpha_1(M_1 M_2)+\alpha_{11}^D(M_1 M_2)]\,
    A([\bar q_s u][\bar u D])
    \nonumber\\[0.2cm]
   &&\mbox{}    + \delta_{pu}\,[\alpha_2(M_1 M_2)+\alpha_{12}^D(M_1 M_2)]\,
    A([\bar q_s D][\bar u u])
    \nonumber\\[0.2cm]
   &&\mbox{}+ [\alpha_3^p(M_1 M_2)+\alpha_3^{\prime\,pD}(M_1 M_2)]\,
    \sum_{q=u,d,s} A([\bar q_s D][\bar q q])
    \nonumber\\[-0.2cm]
   &&\mbox{}    + [\alpha_4^p(M_1 M_2)+\alpha_4^{\prime\,pD}(M_1 M_2)]\,
    \sum_{q=u,d,s} A([\bar q_s q][\bar q D])
    \nonumber\\[-0.2cm]
   &&\mbox{}+ \alpha_{3,\rm EW}^p(M_1 M_2)\,\sum_{q=u,d,s}\frac32\,e_q\,
    A([\bar q_s D][\bar q q])
    + \alpha_{4,\rm EW}^p(M_1 M_2)\,\sum_{q=u,d,s}\frac32\,e_q\,
    A([\bar q_s q][\bar q D])
    \nonumber\\[-0.2cm]
   &&\mbox{}+ \sum_{q=d,s} \alpha_{3q}^p(M_1 M_2)\,A([\bar q_s D][\bar q q])
    + \sum_{q=d,s} \alpha_{4q}^p(M_1 M_2) \,A([\bar q_s q][\bar q D]).
\end{eqnarray}
The new contributions are encoded in $\alpha_{11,12}^D$
(charged Higgs effects), $\alpha_{3,4}^{\prime\,pD}$ (mirror
QCD penguins) and $\alpha_{3q,4q}^p$ (neutral Higgs effects),
as well as modifications of the standard QCD penguin amplitudes
$\alpha_{3,4}^p$. A similar generalization applies to the
annihilation amplitudes. Our aim is to compare the new coefficients
to those present in the SM for PP, PV, VP, VV (P pseudoscalar, V vector
meson) final states. Note that for VV, (\ref{b2}) and (\ref{alphaidef})
apply separately to each of the three independent helicity amplitudes
$h=0,-,+$, but the helicity label is suppressed in our notation.

In (\ref{alphaidef}) $A([\bar q_{M_1} q_{M_1}][\bar q_{M_2} q_{M_2}])$
refers to a product of decay constant, form factor and other factors
\cite{Beneke:2003zv,Beneke:2006hg}, and the arguments indicate the
flavour content of the final state mesons $M_1 M_2$. Since $V\pm A$ and
$S\pm P$ operators contribute differently to pseudoscalar and vector final
states we next write\footnote{In the following we drop the
superscript ``$D$'' (referring to $b\to D$ transitions)
on the amplitude parameters and Wilson coefficients
of the mirror penguin contributions.}
\bea
\alpha_3^{\prime \,p}(M_1 M_2) &=& \left\{
    \begin{array}{cl}
    -a_3^{\prime \,p}(M_1 M_2) + a_5^{\prime \,p}(M_1 M_2),
      & \quad \mbox{if~} M_1 M_2=PP, \\
     a_3^{\prime \,p}(M_1 M_2) + a_5^{\prime \,p}(M_1 M_2),
      & \quad \mbox{if~} M_1 M_2=PV, \\
     a_3^{\prime \,p}(M_1 M_2) - a_5^{\prime \,p}(M_1 M_2),
      & \quad \mbox{if~} M_1 M_2=VP, \\
    -a_3^{\prime \,p}(M_1 M_2) - a_5^{\prime \,p}(M_1 M_2),
      & \quad \mbox{if~} M_1 M_2=V^0V^0, \\
    -f^{M_1}_\pm\left(a_3^{\prime \,p}(M_1 M_2) + a_5^{\prime \,p}(M_1 M_2)\right),
      & \quad \mbox{if~} M_1 M_2=V^{\pm}V^{\pm},
    \end{array}\right.
\nonumber\\
\alpha_4^{\prime \,p}(M_1 M_2) &=& \left\{
    \begin{array}{cl}
    -a_4^{\prime \,p}(M_1 M_2) - r_{\chi}^{M_2}\,a_6^{\prime \,p}(M_1 M_2),
      & \quad \mbox{if~} M_1 M_2=PP, \\
     a_4^{\prime \,p}(M_1 M_2) + r_{\chi}^{M_2}\,a_6^{\prime \,p}(M_1 M_2),
      & \quad \mbox{if~} M_1 M_2=PV, \\
     a_4^{\prime \,p}(M_1 M_2) - r_{\chi}^{M_2}\,a_6^{\prime \,p}(M_1 M_2),
      & \quad \mbox{if~} M_1 M_2=VP, \\
    -a_4^{\prime \,p}(M_1 M_2) + r_{\chi}^{M_2}\,a_6^{\prime \,p}(M_1 M_2),
      & \quad \mbox{if~} M_1 M_2=V^0V^0, \\
    f^{M_1}_\pm\left(-a_4^{\prime \,p}(M_1 M_2) + r_{\chi}^{M_2}\,a_6^{\prime \,p}(M_1 M_2)\right),
      & \quad \mbox{if~} M_1 M_2=V^{\pm}V^{\pm},
    \end{array}\right.
\nonumber\\
\alpha^{p}_{3q}(M_1M_2)&=&
\frac{r_{\chi}^{M_2}}{2}  \left\{ \begin{array}{rl}
   a^{p}_{13q}(M_1M_2), & \quad \mbox{if~} M_1M_2=PP,\,VP, \\
  -a^{p}_{13q}(M_1M_2), & \quad \mbox{if~} M_1M_2=PV,\,V^0V^0, \\
  -f^{M_1}_\pm a^{p}_{13q}(M_1M_2), & \quad \mbox{if~} M_1M_2=V^{\pm}V^{\pm},
\end{array} \right.
\nonumber\\
\alpha^{p}_{4q}(M_1M_2)&=& \frac{1}{2} \left\{
\begin{array}{rl}
  - a^{p}_{14q}(M_1M_2), & \quad \mbox{if~} M_1M_2=PP,\,PV, \\
    a^{p}_{14q}(M_1M_2), & \quad \mbox{if~} M_1M_2=VP,\,V^0V^0, \\
    f^{M_1}_\pm a^{p}_{14q}(M_1M_2), & \quad \mbox{if~} M_1M_2=V^{\pm}V^{\pm};
\end{array}
\right.
\label{b9}
\eea
the same relations as the last two hold between $\alpha_{11}^D$
and $a_{11D}$, and between $\alpha_{12}^D$
and $a_{12D}$, respectively. We denote by
$f^{M_1}_\pm=F_{\mp}^{B\rightarrow M_1}(0)/F_{\pm}^{B\rightarrow M_1}(0)$
a ratio of form factors, such that
$f^{M_1}_+\sim m_B/\Lambda_{\rm QCD}$
and $f^{M_1}_-\sim\Lambda_{\rm QCD}/m_B$ in the heavy-quark
limit \cite{Beneke:2006hg}. It follows that for the transverse helicity
amplitudes of $\bar B\to VV$ decay modes the contributions from the
new operators obey a different
hierarchy in the heavy-quark limit. While in the SM
\be
\label{b11}
\bar{\cal A}_0:\bar{\cal A}_-:\bar{\cal A}_+ =
1:\frac{\Lambda_{\rm QCD}}{m_b} :\frac{\Lambda_{\rm QCD}^2}{m_b^2}
\ee
(up to certain electromagnetic effects \cite{Beneke:2005we}), the
Higgs contributions to the amplitude satisfy
\be
\label{b12}
\bar{\cal A}_0:\bar{\cal A}_-:\bar{\cal A}_+ =
1:\frac{\Lambda_{\rm QCD}^2}{m_b^2} :\frac{\Lambda_{\rm QCD}}{m_b}.
\ee
This effect, noted first in \cite{Das:2004hq}, is interesting, since it
increases the sensitivity of certain polarization observables to the new
short-distance coefficients by a factor $f^{M_1}_+\approx 10$. On the other
hand, the absence of tensor operators implies that the formal
dominance of the longitudinal amplitude is preserved by the Higgs
contributions.

In QCDF the $a_{iq}^{(\prime)\,p}$ coefficients introduced in (\ref{b9})
can be written at next-to-leading order (NLO) in the form
\begin{eqnarray}
   a_{iq}^{(\prime)\,p}(M_1 M_2) &=&
  \left( C_i^{(\prime)\,q} + \frac{C_{i\pm 1}^{(\prime)\,q}}{N_c} \right)
   N_i^{(\prime)}(M_2) \nonumber\\
  && + \,\frac{C_{i\pm 1}^{(\prime)\,q}}{N_c}\,\frac{C_F\alpha_s}{4\pi}
   \left[ V_i^{(\prime)}(M_2) +
   \frac{4\pi^2}{N_c}\,H_i^{(\prime)}(M_1 M_2) \right]
   + P_i^{(\prime)\,p}(M_2) ,
\label{b13}
\end{eqnarray}
where the upper (lower) signs apply when $i$ is odd (even).
The quantities $N_i^{(\prime)}(M_2)$, $V_i^{(\prime)}(M_2)$,
$H_i^{(\prime)}(M_1M_2)$, $P_i^{(\prime)\,p}(M_2)$ stand,
respectively, for the tree-level result (``naive factorization''), the
1-loop vertex correction, spectator scattering, and the
penguin diagrams.

The leading-order (naive factorization) term in (\ref{b13}) is simply
a combination of short-distance coefficients, except for cases where a
vector meson couples to a scalar current, where it is zero. This is
summarized by
\bea
\label{b14}
N_i^{(\prime)}(M_2)=\left\{
\begin{array}{ll}
1, & \mbox{  for } i=3,4,5,12,14q, \\
1, & \mbox{  for } i=6,11,13q \,\mbox{ and }M_2=P, \\
0, & \mbox{  for } i=6,11,13q \,\mbox{ and }M_2=V.
\end{array} \right.
\eea
The NLO coefficients in (\ref{b13}) can mostly be expressed in terms
of those already known from the SM operators
\cite{Beneke:1999br,Beneke:2001ev,Beneke:2003zv,Beneke:2006hg}. For
the mirror QCD penguin operators, we find that they are almost
identical to the SM QCD penguins, that is $V_i^{\prime}(M_2)=V_i(M_2)$,
$H_i^{\prime}(M_1M_2)=H_i(M_1M_2)$ for $i=3\ldots 6$. For the
penguin contribution  $P_{4,6}^{\prime\,p}(M_2)$ one replaces
$C_i\to C_i^\prime$ in the SM expression and then adds the term
$\delta P_{4,6}^{\prime\,p}(M_2)$ from the scalar operators given
in (\ref{b16b}) below.
For the scalar operators, we set $C_{12}^D$ and $C_{14}^q$ to zero
(see section~\ref{weak}) and, using the Fierz symmetry of the NDR
renormalization scheme for the scalar operators~\cite{Buras:2000if},
obtain
\bea
&&
a_{11D}(M_1 M_2) = C_{11}^D\,N_{11},
\nonumber\\
&&
a_{12D}(M_1 M_2) =\frac{C_{11}^{D}}{N_c} + \,\frac{C_{11}^{D}}{N_c}\,
\frac{C_F\alpha_s}{4\pi}
   \left[ V_5(M_2) +
   \frac{4\pi^2}{N_c}\,H_5(M_1 M_2) \right],
\eea
\bea
&&
a_{13q}^p(M_1 M_2) = C_{13}^q\,N_{13q},
\nonumber\\
&&
a_{14q}^p(M_1 M_2) =\frac{C_{13}^{q}}{N_c} + \,\frac{C_{13}^{q}}{N_c}\,
\frac{C_F\alpha_s}{4\pi}
   \left[ V_5(M_2) +
   \frac{4\pi^2}{N_c}\,H_5(M_1 M_2) \right],
\quad
\label{ascalar}
\eea
where $V_5(M_2)$, $H_5(M_1 M_2)$ can be found in \cite{Beneke:2003zv}
for PP, PV, and VP final states and in \cite{Beneke:2006hg} for VV.
(Although not used in the following, since we set $C_{12}^D$ and
$C_{14}^q$ to zero, we note that similarly $V_{11}(M_2) =V_{13}(M_2)
=V_{6}(M_2)$, and $H_{11}(M_1 M_2)=H_{13}(M_1 M_2)=H_{6}(M_1 M_2)$.)

\begin{figure}[t]
  \centering
  \includegraphics[height=2.0cm]{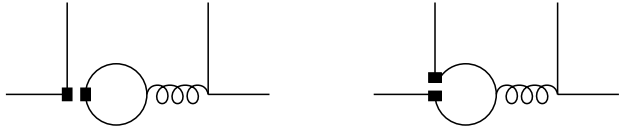}
  \caption{Penguin contractions. Due to colour only the second diagram
contributes to insertions of $Q_{11}^p$, $Q_{13}^q$.}
  \label{figd3}
\end{figure}

There are no penguin contributions to (\ref{ascalar}). However, as
discussed above, the insertion of scalar operators into the penguin
diagrams shown in figure~\ref{figd3} modifies the evolution of the
(mirror) QCD penguin operators. Accordingly, it also contributes
to the penguin terms $P_{4,6}^{(\prime)\,p}(M_2)$ in
$a_{4,6}^{(\prime)\,p}(M_1 M_2)$. The correction terms are proportional
to the coefficient functions of the scalar operators and read
\bea
\delta P_{4}^p(M_1M_2) &=&\frac{C_F \alpha_s}{4 \pi N_c}
\left(-\frac{1}{2}\right) C_{13}^{b} \left[\frac{4}{3} \log
\frac{m_b}{\mu}-G_{M_2}^{f}(1) \right],
\nonumber\\
\delta P_{6}^p(M_1M_2) &=&\frac{C_F \alpha_s}{4 \pi N_c}
\left(-\frac{1}{2}\right) C_{13}^{b} \left[N_6(M_2)\,\frac{4}{3} \log
\frac{m_b}{\mu}  -\hat G_{M_2}^{f}(1) \right],
\label{b16}
\eea
where $G_{M_2}^f(s)$ equals $G_{M_2}(s)$ \cite{Beneke:2003zv} for
$M_1 M_2=PP,PV,VP,V^0 V^0$, and  $G_{M_2}^\pm(s)$ \cite{Beneke:2006hg} for
$V^{\pm} V^{\pm}$ with
\be
\label{b26}
G^+_{V_2}(s) = \int_0^1 dy \,\phi_{a2}(y)
\,G(s-i \epsilon, 1-y),
\ee
while $\hat G_{M_2}^f(s)$ equals $\hat G_{M_2}(s)$ for
$M_1 M_2=PP,PV,VP,V^0 V^0$, and is zero for
$V^{\pm} V^{\pm}$. Similarly, for the mirror penguin coefficients
\bea\label{b16b}\nn
\delta P_{4}^{\prime \,p}(M_1M_2) &=& \frac{C_F \alpha_s}{4 \pi N_c}
\left(-\frac{1}{2}\right) \bigg\{
C_{13}^{D} \left[\frac{4}{3} \log
\frac{m_b}{\mu}- G_{M_2}^f(0) \right] \\
&&\hspace*{0cm}+\,C_{11}^{D} \left[\frac{4}{3}
\log \frac{m_b}{\mu}-G_{M_2}^f(s_p) \right] \bigg\},
\nonumber\\
\delta P_{6}^{\prime \,p}(M_1M_2) &=& \frac{C_F \alpha_s}{4 \pi N_c}
\left(-\frac{1}{2}\right) \bigg\{
C_{13}^{D} \left[N_6(M_2)\,\frac{4}{3} \log
\frac{m_b}{\mu}- \hat G_{M_2}^f(0) \right]
\nonumber\\
&&\hspace*{0cm}+\,C_{11}^{D} \left[N_6(M_2)\,\frac{4}{3}
\log \frac{m_b}{\mu}-\hat G_{M_2}^f(s_p) \right] \bigg\},
\eea
where $s_u=0$, $s_c=(m_c/m_b)^2$, and
now  $G_{M_2}^f(s)$ equals $G_{M_2}^\mp(s)$ for
$V^{\pm} V^{\pm}$ and else as above. Note that the explicit scale dependence
in $\delta P_{4,6}^{(\prime)\,p}(M_2)$ cancels the extra scale dependence
of the (mirror) QCD penguin coefficients at LL
accuracy. At this point we should mention that the constant terms in the
real part of the NLO matrix elements should strictly speaking only be
considered at the NLL order. At this order, our calculation is, however,
incomplete, since we do not consider the 1-loop QCD correction to the
initial condition of the scalar operators, and the 2-loop mixing into
the penguin operators, as well as the small contributions from $C_{12}^D$
and $C_{14}^q$. Since we do not need precise results for the NLO terms,
as will be seen below, the present approximation is adequate for our
purpose. However, the complete NLO results for the
matrix elements of scalar and mirror penguin operators given above
might be of more general interest.

We also calculated the weak annihilation terms ${\cal{T}}_B^p$ originating
from the scalar operators. In some cases the annihilation amplitude
can be as large as the corresponding $\alpha_i$ amplitude. Since
no precise estimates are needed below, we do not discuss the annihilation
amplitudes further.\footnote{We use this occasion to point out
the following corrections to \cite{Beneke:2006hg}: The overall sign on
the right-hand side of (A.15) [eq.(63) in the arXiv version] must be
minus. Furthermore, the expression for $A_3^{i,0}$ [$A_3^{f,0}$]
in (A.20) [eq.(68)] must contain $r_\chi^{V_1}-r_\chi^{V_2}$
[$r_\chi^{V_1}+r_\chi^{V_2}$] rather than the opposite relative
sign \cite{Cheng:2008gxa}.
(However, the unsimplified expressions in (A.18) [eq.(66)] are given
correctly.)}

\section{Non-leptonic decays}
\label{pheno}

We are now ready to discuss the question whether there are observable
effects on non-leptonic, charmless decays due to Higgs exchange in the
MSSM with large $\tan\beta$. To this end, we compare the new amplitudes
to those present in the SM. The essential features can be deduced from
(\ref{alphaidef}).
\begin{itemize}
\item Charged Higgs exchange ($\alpha_{11,12}^D$) contributes directly to
tree-dominated decays (such as $B\to \pi\pi,\pi\rho,\rho\rho$), but must
compete with the sizeable SM tree amplitudes $\alpha_{1,2}$. However, since
$\alpha_{11,12}^D \propto \bar m_D$, only the case of $b\to s\bar u u$
transitions is of interest. But there are no tree-dominated decays
of this type, since for $D=s$ the tree amplitudes are doubly
CKM-suppressed, $\lambda^{(s)}_u\ll \lambda^{(s)}_c$.
\item The effects from the mirror QCD penguin operators
($\alpha_{3,4}^{\prime\,p}$) must compete with the SM penguin amplitudes,
which according to (\ref{pen1}) requires the scalar operator Wilson
coefficients to be of order 1 in general, and of order $0.1$ in case
of the plus-helicity amplitude in $\bar B\to VV$.
\item The direct contribution from the FCNC Higgs couplings
($\alpha_{3q,4q}^p$) is an isospin-violating effect that must compete
only with the small SM electroweak penguins, and is therefore most likely to
lead to an observable effect. Since $\alpha_{3q,4q}^p\propto \bar m_q$,
only the case $q=s$ is of interest. For the case of  $D=s$,
the $b\to s\bar s s$ transition leads to final states with
flavour content $M_1=\bar q_s s$, $M_2=\bar s s$ with $\bar q_s$
the flavour of the $\bar B$ meson spectator antiquark. This singles out
the decay modes $\bar B\to \bar K^{(*)} (\eta^{(\prime)},\phi)$ and
$\bar B_s \to (\eta^{(\prime)},\phi) (\eta^{(\prime)},\phi)$.
For the case of $D=d$, the potentially interesting modes are
 $\bar B\to \bar K^{(*)}  K^{(*)}$ and
$\bar B_s \to  K^{(*)} \phi$. However,
in all these decays it is impossible to extract the EW penguin amplitude,
so the new contributions must in fact be compared to the larger
SM QCD penguins.
\end{itemize}
We now proceed to a more detailed discussion. The numerical amplitude
values given below depend on parameters (quark masses, form factors,
etc.), for which we choose values as given
in \cite{Beneke:2003zv,Beneke:2006hg}, including some updates. Since none
of our conclusions depends on the precise values of these parameters,
we do not list them here. The Wilson coefficients are evaluated at
the scale $\mu=m_b=4.2\,$GeV.

\subsubsection*{\boldmath $B\rightarrow PP, PV$}
\label{bpp}

\begin{table}[tp]
  \centering
  \begin{tabular}{|c|c|c|c|}
    \hline
& $\bar{K} \eta_s$ & $\bar{K}^* \eta_s$ & $\bar{K} \phi$ \\ \hline
  $\alpha_{1}$          &$0.966+0.021 i~[\pi \bar K]$
&$0.981+0.021 i~[\rho \bar K]\hskip0.2cm$
&$0.973+0.021 i~[\pi \bar K^*]$ \\
  $\alpha_{2}$          &$0.351-0.084 i~[\bar K \pi]$&
$0.260-0.084 i~[\bar K^* \pi]$&
$0.323-0.084 i~[\bar K \rho]\hskip0.2cm$ \\ \hline
  $\alpha_{11}^s$       &       $-0.059~[\pi \bar K]$
&        $-0.059~[\rho \bar K]$&
$0~[\pi \bar K^*]$ \\
  $\alpha_{12}^s$       &$0.003+0.003 i~[\bar K \pi]$&
$-0.006-0.003 i~[\bar K^*\pi]$
&$0.004+0.003 i~[\bar K \rho]$ \\ \hline
  $\alpha^{u}_{3}$           &$-0.0013+0.0046 i$& $0.0027+0.0046 i$&
$0.0006-0.0005 i$ \\
  $\alpha^{u}_{4}$           &  $-0.095-0.040 i$& $0.038+0.008 i$&
$-0.031-0.017 i$ \\ \hline
  $\delta P^{u}_{4}$
&$1.4\cdot 10^{-5}$&$1.4\cdot 10^{-5}$&$\phantom{-}1.4\cdot 10^{-5}$\\
  $\delta P^{u}_{6}$
&$1.4\cdot 10^{-5}$&$1.4\cdot 10^{-5}$&$-1.5\cdot 10^{-5}$\\ \hline
  $\alpha^{\prime\,u}_{3}$   & $7.8\cdot10^{-5}-0.0001 i$&$2.8\cdot10^{-5}+0.0001 i$ &
$(1.4-1.3i)\cdot 10^{-5}$ \\
  $\alpha^{\prime\,u}_{4}$   & $\phantom{-}0.0035+0.0015 i$&
$\phantom{-}0.0011+0.0003i$&$-0.0013-0.0006i$ \\ \hline
  $\delta P^{\prime\,u}_{4}$ &$-0.0005-0.0007 i$&$-0.0005-0.0007
i$&$-0.0005-0.0007 i$\\
  $\delta P^{\prime\,u}_{6}$ &$-0.0006-0.0007 i$&$-0.0006-0.0007 i$&
$-0.0003$\\ \hline
  $\alpha^{u}_{3,\rm EW}$        &$-0.0089-0.0002 i$&$-0.0091-0.0002
i$&$-0.0082-0.0001i$ \\
  $\alpha^{u}_{4,\rm EW}$        &$-0.0016+0.0006 i$&$-0.0025+0.0008
i$&$-0.0024+0.0007i$ \\ \hline
  $\alpha^{u}_{3s}$          &         $0.00078$&         $0.00078$&
$0$ \\
  $\alpha^{u}_{4s}$
&$(-6.3-3.7i)\cdot 10^{-5}$&$(9.7+3.7i)\cdot 10^{-5}$&
 $(-6.3-3.7i )\cdot 10^{-5}$ \\ \hline
  \end{tabular}
  \caption{Numerical results for the $\alpha_i$ coefficients of
  some representative $\Delta S=1$ decay channels.
  The value of $\alpha^{u}_{4}$ corresponds to the SM contribution
  only. The final states $\bar K\eta _s$, $\bar K^*\eta_s$,
  $\bar K\phi$ do not receive tree-amplitude contributions. For $\alpha_{1,2}$
  and $\alpha_{11,12}^s$, we therefore provide results for the final states
  in square brackets.}\label{tab10}
\end{table}

In table \ref{tab10} we show the numerical results of the
$\alpha_i$ amplitude coefficients defined in (\ref{alphaidef}) for
the decay modes  $\bar B\to \bar K \eta, \bar K^* \eta, \bar K \phi$.
($\eta_s$ in the table refers to the strange component of $\eta$,
see \cite{Beneke:2002jn}.) To evaluate the Higgs contributions
we assume the largest values of the coefficient functions allowed
by the constraints from leptonic decays derived in section~\ref{Bsmumu},
in detail: $C_{11}^s(m_b)= -0.08$, $C_{13}^s(m_b)= 0.001$,
$C_{13}^b(m_b)= 0.05$.

Among the Higgs penguin amplitudes $\alpha_{3q}^p$ is the larger of
$\alpha_{3q,4q}^p$, since
$\alpha_{4q}^p$ is colour-suppressed and is further reduced by the radiative
correction given in (\ref{ascalar}). However, the strong constraint
on $C_{13}^s$ renders  $\alpha_{3q}^p$ always negligible, in particular
as it should be compared to the QCD penguin amplitude $\alpha_4^p$ rather
than the electroweak penguin. This remains true for VP amplitudes despite
the fact that the SM penguin amplitude is smaller for these final states,
and for PV amplitudes, where  $\alpha_{3q}^p$ vanishes.

\begin{table}
  \centering
  \begin{tabular}{|c|c|c|c|}
    \hline
& $(\bar K^{*}\phi)^{00}$ & $(\bar K^{*}\phi)^{--}$ &
$(\bar K^{*}\phi)^{++}$ \\ \hline
  $\alpha_{1}~[\rho \bar K^*]$    & $0.987 + 0.021 i$
& $\phantom{-}1.101 + 0.041 i$&
$1.018$ \\
  $\alpha_{2}~[\bar K^* \rho]$    & $0.240 - 0.084 i$&$-0.173 - 0.169 i$&
$0.170$ \\ \hline
  $\alpha_{11}^s~[\rho \bar K^*]$ &               $0$&               $0$&
$0$ \\
  $\alpha_{12}^s~[\bar K^* \rho]$ &$-0.007 - 0.003 i$&   $-0.002$&$-0.247
- 0.068 i$ \\ \hline
  $\alpha^{u}_{3}$           & $0.0001-0.0005 i$&$-0.0023-0.0010 i$&
$-0.0035$ \\
  $\alpha^{u}_{4}$           &  $-0.026-0.015 i$&  $-0.044-0.017 i$&
$-0.031$ \\ \hline
  $\delta P^{u}_{4}$
&$\phantom{-}1.4\cdot 10^{-5}$&$0.7\cdot 10^{-5}$&$2.2\cdot 10^{-5}$ \\
  $\delta P^{u}_{6}$         &$-1.5\cdot 10^{-5}$&              $0$&
$0$ \\ \hline
  $\alpha^{\prime\,u}_{3}$   &$(-0.2+1.3i)\cdot10^{-5}$&$(5.1+2.3i)\cdot10^{-6}$&
0.0010  \\
  $\alpha^{\prime\,u}_{4}$   & $\phantom{-}0.0011+0.0006
i$&$0.0001+5.5\cdot 10^{-5}i$& $\phantom{-}0.0173 + 0.0074 i$ \\ \hline
  $\delta P^{\prime\,u}_{4}$ &$-0.0005-0.0007
i$&$-0.0004-0.0007i$&$-0.0007-0.0007 i$ \\
  $\delta P^{\prime\,u}_{6}$ &         $-0.0003$&               $0$&
$0$ \\ \hline
  $\alpha^{u}_{3,\rm EW}$        &$-0.0084-0.0001 i$& $0.0044-0.0003 i$&
$-0.009$ \\
  $\alpha^{u}_{4,\rm EW}$        &$-0.0017+0.0007 i$& $0.0015+0.0014 i$&
$-0.0015$ \\ \hline
  $\alpha^{u}_{3s}$          &               $0$&               $0$&
$0$ \\
  $\alpha^{u}_{4s}$
&$(9.7+3.7i)\cdot 10^{-5}$&$2.0\cdot 10^{-5}$& $0.0031+0.0008 i$
\\ \hline
  \end{tabular}
  \caption{
Numerical results for the $\alpha_i$ coefficients pertaining to the
three helicity amplitudes of $\bar B\to VV$ decays.
The value of $\alpha^{u}_{4}$ corresponds to the SM contribution only.
To compare the
absolute values of the helicity amplitudes the numbers for
the $(00,--,++)$ parameters
must be multiplied by $A_{K^*\phi}=\frac{i G_{\rm F}}{\sqrt{2}}m_B
f_{\phi} \,(m_B A_0^{B\to K^*},
m_{\phi} F_-^{B\to K^*}, m_{\phi} F_+^{B\to K^*})$. The
estimates above use $F_+^{B\to V_1}=0.06$ in order to compare with the
maximal SM $++$ amplitude.  The final state
$\bar K^*\phi$ does not receive tree contributions. For $\alpha_{1,2}$
and $\alpha_{11,12}^s$, we therefore provide results for the final states
in square brackets.}\label{tab16}
\end{table}

\subsubsection*{\boldmath$B\rightarrow VV$}
\label{bvv}

The effect of Higgs exchange is also negligible in case of the
longitudinal amplitude in $\bar B\to VV$ decays, since it follows
the same pattern as for the PV decays with $M_2=V$.
Due to the inverted hierarchy of the transverse polarization
amplitudes, see~(\ref{b12}), the minus-helicity amplitude is suppressed,
while the plus-helicity amplitude is enhanced by a factor of
$m_b/\Lambda_{\rm QCD}$ relative to the SM. To compare
the Higgs contributions to the plus-helicity amplitude in the SM, in
table~\ref{tab16} we show the $\alpha_i$ coefficients
assuming $F_+^{B\to V_1}=0.06$, which is the upper limit allowed
in~\cite{Beneke:2006hg}. It is evident that
the magnitude of the Higgs-induced $\alpha_i$ coefficients is
now larger for the plus amplitude
than for PP, PV final states and the other polarization
amplitudes. In fact, $\alpha_{3s}^u$ would now be comparable to
the SM penguin amplitude, if it were not annihilated by the
projection on the vector meson at tree level, see~(\ref{b14}).
Thus, among amplitudes with the same flavour topology, we find
that only $\alpha_{12}^s$ is larger than the corresponding SM
colour-suppressed tree amplitude $\alpha_2$. The mirror QCD 
penguin amplitude $\alpha_4^{\prime\,u}$ amounts to a substantial 
fraction of the standard penguin amplitude that may reach one 
if $F_+^{B\to V_1}$ is smaller than the assumed value. This would affect
the azimuthal angular distribution of $\Delta S=1$ decays; in
practice, however, the effect is unobservable. Not only is the
amplitude very small in absolute terms, but the tree amplitudes are
also subleading to the penguin amplitudes in
$\Delta S=1$ decays.

\vspace*{0.4cm}
\noindent
It is straightforward to compute branching fractions,
CP asymmetries and polarization observables including the Higgs-exchange
contributions. However, since the $\alpha_i$ parameters discussed
above form the basic constituents of observables, it follows
that any modification of the SM predictions will be invisible within
theoretical uncertainties.

\section{Conclusion}
\label{concl}

Motivated by the interest in the minimally flavour-violating
MSSM with large $\tb$ owing to its potentially large impact on
leptonic $B$ decays, we analyzed non-leptonic $B$ decays
in this model. The hadronic and leptonic flavour-changing
interactions are closely related, which allows us to
translate the present limit on the $B_s\to \mu^+\mu^-$ branching
fraction, and the observation of $B^+\to \tau^+ \nu_\tau$ into
a constraint on the Wilson coefficients of the relevant
scalar four-quark operators.
We then calculated the matrix elements of scalar operators and mirror
QCD penguin operators at next-to-leading order in the framework
of QCD factorization and find that the limits on leptonic
$B$ decay branching fractions exclude any
visible effects in hadronic decays, but for an academic exception:
the positive-helicity amplitude of $\bar B\to VV$ may receive
order one modifications relative to the SM, but this amplitude is
too small to be detected at present or planned $B$ factories.
\vspace{1em}

\noindent
\subsubsection*{Acknowledgement}
We are grateful to J.~Rohrer and De-shan Yang for many discussions and
comparisons on QCD factorization results for $B \to VV$ decays. M.~B.
thanks the CERN Theory group for hospitality. This
work is supported by the DFG
Sonder\-forschungs\-bereich/Transregio~9 ``Computergest\"utzte
Theoretische Teilchenphysik''. X.~Q.~Li acknowledges support from
the Alexander-von-Humboldt Foundation.

\end{document}